\documentclass[12pt]{iopart}
\usepackage{iopams}
\usepackage{graphicx,bbm,mathrsfs,pstricks,psfrag}  
\newcommand{\E}{\mathcal{E}_\epsilon}
\newcommand{\emax}{\epsilon_{\sf max}}

\newtheorem{criterion}{Criterion}
\begin{document}

\title{Detecting quantum non-Gaussianity of noisy Schr\"odinger cat states}

\author{ Mattia L. Palma$^{1,2}$, Jimmy Stammers$^{2}$, Marco G. Genoni$^2$, Tommaso Tufarelli$^2$, Stefano Olivares$^1$, M. S. Kim$^2$ and Matteo G. A. Paris$^1$}
\address{$^1$ Dipartimento di Fisica dell'Universit\`a degli Studi
di Milano, I-20133 Milano, Italy.}
\address{$^2$ QOLS, Blackett Laboratory, Imperial College London, London SW7 2BW, UK.}
\date{\today}
\ead{m.genoni@imperial.ac.uk}
\begin{abstract} 
Highly quantum non-linear interactions between different bosonic modes lead to the generation of quantum non-Gaussian states, {\em i.e.} states that cannot be written as mixtures of Gaussian states. A paradigmatic example is given by Schr\"odinger's cat states, that is coherent superpositions of coherent states with opposite amplitude. We here consider a novel quantum non-Gaussianity criterion recently proposed in the literature and prove its effectiveness on Schr\"odinger cat states evolving in a lossy bosonic channel. We prove that quantum non-Gaussianity can be effectively detected for high values of losses and for large coherent amplitudes of the cat states.
\end{abstract}

\maketitle

\section{Introduction}
The achievement and control of optical nonlinearities at the quantum level (QNL) is arguably one of the central goals of modern quantum optics\footnote{In this manuscript, we use the term `nonlinearity' to indicate any process that cannot be realized with Hamiltonians that are second degree polynomials in the bosonic operators. More technically, a nonlinear process cannot be obtained as a convex combination of Gaussian operations.}. Highly Hamiltonian nonlinear processes such as Kerr interactions, or conditional operations as photon addition/subtraction and Fock state generation \cite{tutorialMSK} have indeed proven to be powerful tools to investigate and exploit the quantum nature of the electromagnetic field. In light of this, techniques that reliably verify the succesful experimental implementation of QNL are highly desirable. Closely linked to the concept of optical nonlinearity is that of the set of {\it Gaussian states} \cite{AOP}, which can be seen as the collection of states that can be obtained by applying quadratic Hamiltonians to thermal states of radiation. It is easy to realize that, without the use of QNL, one is invariably limited to the preparation of Gaussian states and their convex combinations. Conversely, the successful detection of a state that cannot be written in this form, a {\it Quantum non-Gaussian state}, can only be explained by the presence of QNL during the preparation stage. The detection and characterization of quantum non-Gaussianity (QNG) thus acquires fundamental importance in the study of continuous variable quantum states. The literature presents a number of methods to detect non-classical states \cite{glauber}, defined as states that cannot be written as mixtures of coherent states, or to quantify the deviation of a quantum state from a Gaussian \cite{geno1,geno2,geno3,barbieri}. However these methods are respectively not able to discriminate between quantum non-Gaussian states and squeezed states, or not suitable to distinguish between quantum non-Gaussian states and mixtures of Gaussian states. In fact excluding the case of states with negative Wigner function, which are certainly quantum non-Gaussian, no general method is known to distinguish between the two sets. This state of knowledge triggered the development of sufficient methods to detect quantum non-Gaussianity in noisy setups, where no negativity of the Wigner function can be observed \cite{filip1,qnonGPRA}, allowing to witness the succesful implementation of QNL processes despite the high levels of noise \cite{filip2,filip3}. In this paper we apply the method introduced in \cite{qnonGPRA}, to investigate QNG of {\it Schr\"odinger cat states} \cite{cats1,cats2,cats3} undergoing severe optical loss, such that its Wigner function becomes everywhere positive. Focusing on the so-called {\it odd} and {\it even} cat states, we find that QNG can be witnessed for any value of the model parameters in the former case, and for a significant but finite range of parameters in the latter. In what follows, we start by briefly summarizing the results of Ref.~\cite{qnonGPRA} and the basics of the employed physical model, and then use these tools to carry out a detailed analysis of the problem of interest.  
\section{Quantum non-Gaussianity criteria}
We here review QNG criteria based on the Wigner function which have been proposed in \cite{qnonGPRA}. We will restrict here to single-mode systems, descrbed by bosonic operators satisfying the commutation relation $[a, a^\dag] = \mathbbm{1}$. Any single mode quantum state
$\varrho$ can be equivalently described by its characteristic function or its Wigner function, defined
respectively as 
\begin{equation}
\chi[\varrho](\gamma) = \Tr[\varrho D(\gamma)],\:\:\: W[\varrho](\alpha) = \int \frac{d^2 \gamma}{\pi^2} e^{\gamma^* \alpha - \gamma \alpha^*} 
\chi[\varrho](\gamma)\:,
\end{equation}
where $D(\gamma) = \exp\{\gamma a^\dag - \gamma^* a \}$ represents the displacement
operator. A quantum state is called Gaussian if and only if its Wigner function is a Gaussian function. \\
The Gaussian convex hull is the set of states
\begin{equation}\label{eq:Ghull}
\mathcal{G} = \left\{ \varrho \in \mathcal{H} \: | \: \varrho = \int d\blambda \: p(\blambda) \: |\psi_{\sf G}(\blambda)\rangle\langle\psi_{\sf G}(\blambda)|  \right\} \:,
\end{equation}
where $\mathcal{H}$ denotes the Hilbert space of continuous-variable quantum states,
$p(\blambda)$ is a proper probability 
distribution and $|\psi_{\sf G}(\blambda)\rangle$ are pure 
Gaussian states. We define a quantum state {\em quantum non-Gaussian} iff it is not possible to express it as a convex mixture of Gaussian
states, that is iff $\varrho \notin \mathcal{G}$. In \cite{qnonGPRA} it is proved that for any $\varrho \in \mathcal{G}$, the following inequality holds
\begin{equation}
W[\varrho](0) \geq \frac 2 \pi e^{-2 \bar{n}(\bar{n}+1)} \:, \label{eq:bound}
\end{equation}
where $\bar{n}=\Tr[\varrho a^\dag a]$. Together with the observation that the set $\mathcal{G}$ is closed under any Gaussian map $\mathcal{E}_{\sf G}$, inequality (\ref{eq:bound}) leads to the following generalized QNG criterion
\begin{criterion}\label{c:criterion}
Given a quantum state $\varrho$ and a Gaussian map $\mathcal{E}_{\sf G}$, define the QNG witness
\begin{equation}
\Delta[\varrho,\mathcal{E}_{\sf G}] = W[\mathcal{E}_{\sf G}(\varrho)](0) - \frac2\pi \exp\{-2 \bar{n}_\mathcal{E}(\bar{n}_\mathcal{E}+1)\}\:, \label{eq:Delta2}
\end{equation}
where $\bar{n}_\mathcal{E} = \Tr[\mathcal{E}_{\sf G}(\varrho) a^\dag a]$.
Then,
\begin{equation}
\exists\, \mathcal{E}_{\sf G}\:\: {\rm s.t.} \:\: \Delta[\varrho,\mathcal{E}_{\sf G}] < 0 
\Rightarrow \: \varrho \notin \mathcal{G}. \nonumber
\end{equation}
\end{criterion}
In the following section we will apply this criterion to Schr\"odinger's cat states evolving in a lossy channel. As additional Gaussian maps $\mathcal{E}_{\sf G}$ we will consider the simplest examples, that is displacement operations $D(\beta)$, squeezing operations 
$S(\xi) = \exp\left\{ \frac12 \xi (a^\dag)^2 - \frac12 \xi^* a^2 \right\}$, and combinations of the two.
\section{Detecting quantum non-Gaussianity of Schr\"odinger's cat states}
\begin{figure}[t!]
\centering
\includegraphics[width=0.45\columnwidth]{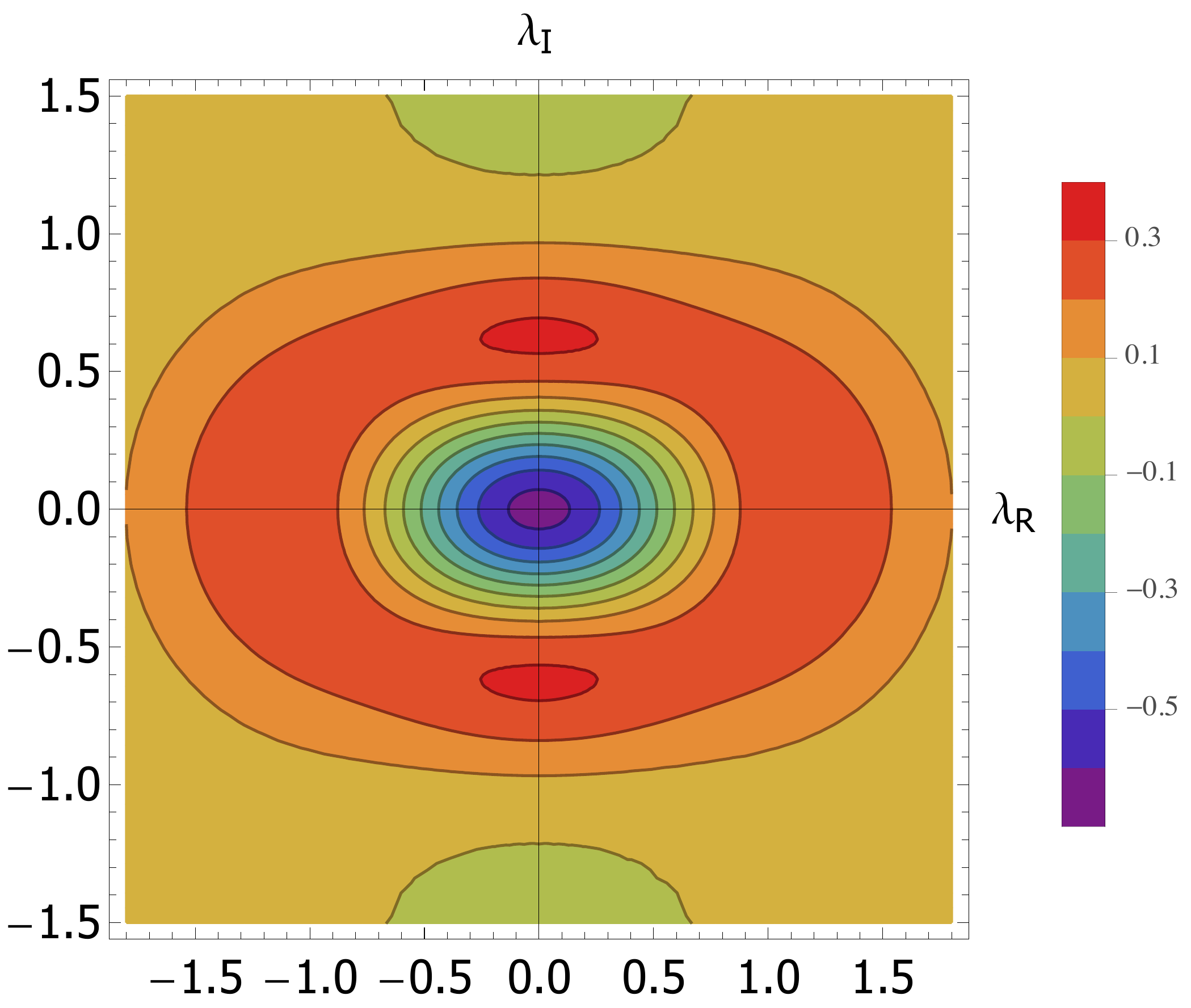}
\includegraphics[width=0.45\columnwidth]{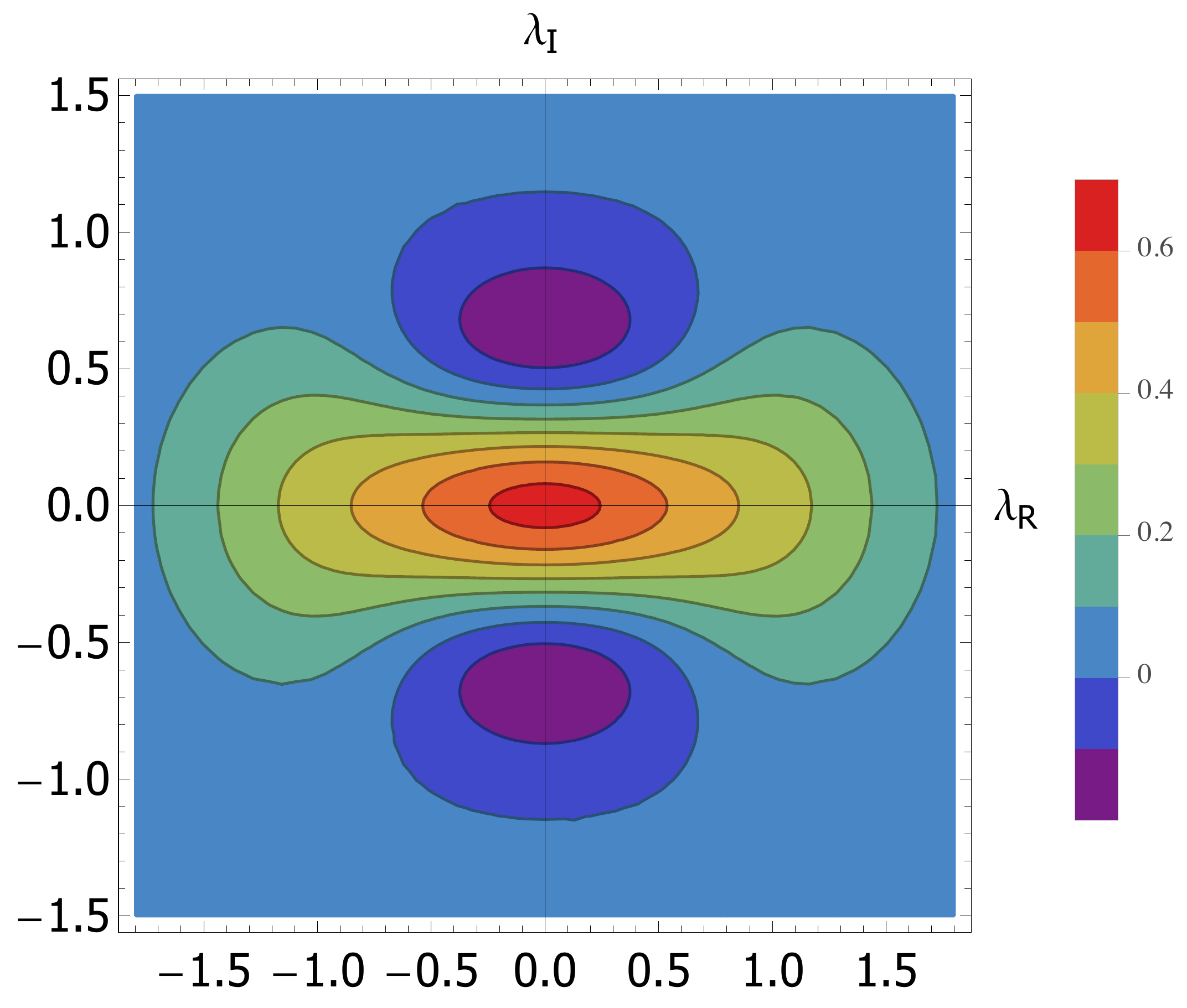}
\caption{Contour plots of the Wigner functions $W[\varrho](\lambda)$ of the odd cat state (left) and even cat state (right) for $\alpha=1$.
\label{f:Wigner}}
\end{figure}
Schr\"odinger's cat states are defined as 
\begin{equation}
|\psi_{\alpha,\xi} \rangle = \frac{|-\alpha\rangle + \xi |\alpha\rangle }{\mathcal{N}}
\end{equation}
where, without losing generality, $\alpha \in \mathbbm{R}$ and $\mathcal{N}=\sqrt{1+\xi^2 + 2\xi e^{-2 \alpha^2}}$ denotes the normalization constant. By considering the parameter $\xi=1$ and $\xi=-1$, one obtains respectively the so-called {\em even} $|\psi_{\sf even}\rangle$ and {\em odd} $|\psi_{\sf odd}\rangle$ cat states. In the following we will restrict our analysis to these particular classes of states, whose Wigner functions are plotted in Fig. \ref{f:Wigner} for $\alpha=1$.
We will consider their evolution in lossy bosonic channel described by the Markovian
master equation 
\begin{equation} \label{eq:ME}
\frac{d\varrho}{dt} = \gamma a \varrho a^\dag - \frac\gamma{2} ( a^\dag a \varrho + \varrho a^\dag a).
\end{equation}
The resulting time evolution is characterized by a single parameter $\epsilon=1-e^{-\gamma t}$ and 
it models both the incoherent loss of photons in a zero temperature environment, and the performances
of detectors having an efficiency parameter $\eta=1-\epsilon$. We will denote the evolved state as
$\mathcal{E}_\epsilon(\varrho_0)$. In the Wigner function picture, the evoution can be analitically solved by 
means of the formula 
\begin{equation}
W[\E(\varrho_0)](\lambda)=\frac{2}{\pi\epsilon}
\int \!\!d^2 \lambda' \: W[\varrho](\lambda^\prime)
\exp\left\{-\frac{2\left| \lambda-\lambda' \sqrt{1-\epsilon}\right|^2}{\epsilon}  \right\} 
\;. \label{eq:WigEv}
\end{equation}
Also the average values of the operators needed to compute the QNG witnesses 
$\Delta[\E(\varrho_0),\mathcal{E}_{\sf G}]$ can be analitically evaluated as
\begin{eqnarray}
\bar{n}_\epsilon &= \Tr[\E(\varrho_0) a^\dag a] = (1-\epsilon) \bar{n}_0 \:,\:\:
\langle a^2 \rangle_\epsilon &= \Tr[\E(\varrho_0) a^2] = (1-\epsilon) \langle a^2 \rangle_0  \:, 
\label{eq:nepsilon} 
\end{eqnarray} 
where for an initial Schr\"odinger's cat state $\varrho_0 = |\psi_{\alpha,\xi}\rangle\langle\psi_{\alpha,\xi}| $ the initial averages read
\begin{equation}
\bar{n}_0 = \frac{\alpha^2 (1+\xi^2-2\xi e^{-2 \alpha^2})}{\mathcal{N}^2} \:, \:\:\:\
\langle a^2 \rangle_0 = \alpha^2 \:.
\label{eq:n0}
\end{equation}
We will focus on large noisy parameters, {\em i.e.} $\epsilon>0.5$ such that no negativity of the Wigner function can be observed \cite{paavola}. In particular we will determine the maximum values for which we observe a violation 
\begin{equation}
\epsilon_{\sf max}[\varrho] = \max\{ \epsilon \: :\exists\mathcal{E}_{\sf G}\: \rm{ s.t.} \ \Delta[\E(\varrho),\mathcal{E}_{\sf G}] \leq 0 \} \:.
\end{equation}
The quantity $\epsilon_{\sf max}$ is a relevant figure of merit to assess
our criterion. In fact, having values of $\epsilon_{\sf max}$ close to unit
corresponds to situations where the criterion is able to detect QNG
even in highly noisy channel or, equivalently, by using highly
inefficient detectors.

\subsection{Odd cat states}
We will start here by considering a odd cat state $|\psi_{\sf odd}\rangle$. The Wigner function of the initial pure state, plotted in Fig. \ref{f:Wigner} (left), is squeezed along the $P$ quadrature and presents a minimum (negative) value at the origin of the phase space. 
Then we can  consider the QNG witness optimized over an additional squeezing operation, {\em i.e.}
\begin{eqnarray}
\Delta_{\sf odd}(s) &= \Delta[\E(|\psi_{\sf odd}\rangle\langle\psi_{\sf odd}|, S(s)] \:.
\end{eqnarray}
The average photon number of the squeezed evolved odd cat state, needed to determine
$\Delta_{\sf odd}(s)$, can be evaluated as
\begin{equation}
\bar{n}^{\sf (odd)} (s) 
= (\mu_s^2 + \nu_s^2) \bar{n}_\epsilon + 2\mu_s\nu_s \langle a^2\rangle_\epsilon + \nu_s^2\:, \label{eq:nodds}
\end{equation}
where $\mu_s=\cosh s$, $\nu_s = \sinh s$, and the values of $\bar{n}_\epsilon$ and $\langle a^2 \rangle_\epsilon$ can be obtained from Eqs. (\ref{eq:nepsilon}) and (\ref{eq:n0}) by setting $\xi=-1$.
We will then look for values of the additional squeezing parameter $s$, such that the criterion \ref{c:criterion} is fulfilled. 
In particular one can then try to optimize over the additional squeezing $s$, for each values of $\alpha$ and $\epsilon$. By exploiting the invariance of the Wigner function in the origin under squeezing operation, this optimization corresponds to the minimization of the average photon number in Eq. (\ref{eq:nodds}).
An analytic solution can be obtained, yielding
\begin{equation}
s_{\sf opt} = - \frac14 \log \frac {1-e^{2\alpha^2} -4 \alpha^2 e^{2\alpha^2}(1-\epsilon)}{1-e^{2\alpha^2}- 4\alpha^2(1-\epsilon)} .
\end{equation}
%
\begin{figure}[h!]
\centering
\includegraphics[width=0.4\columnwidth]{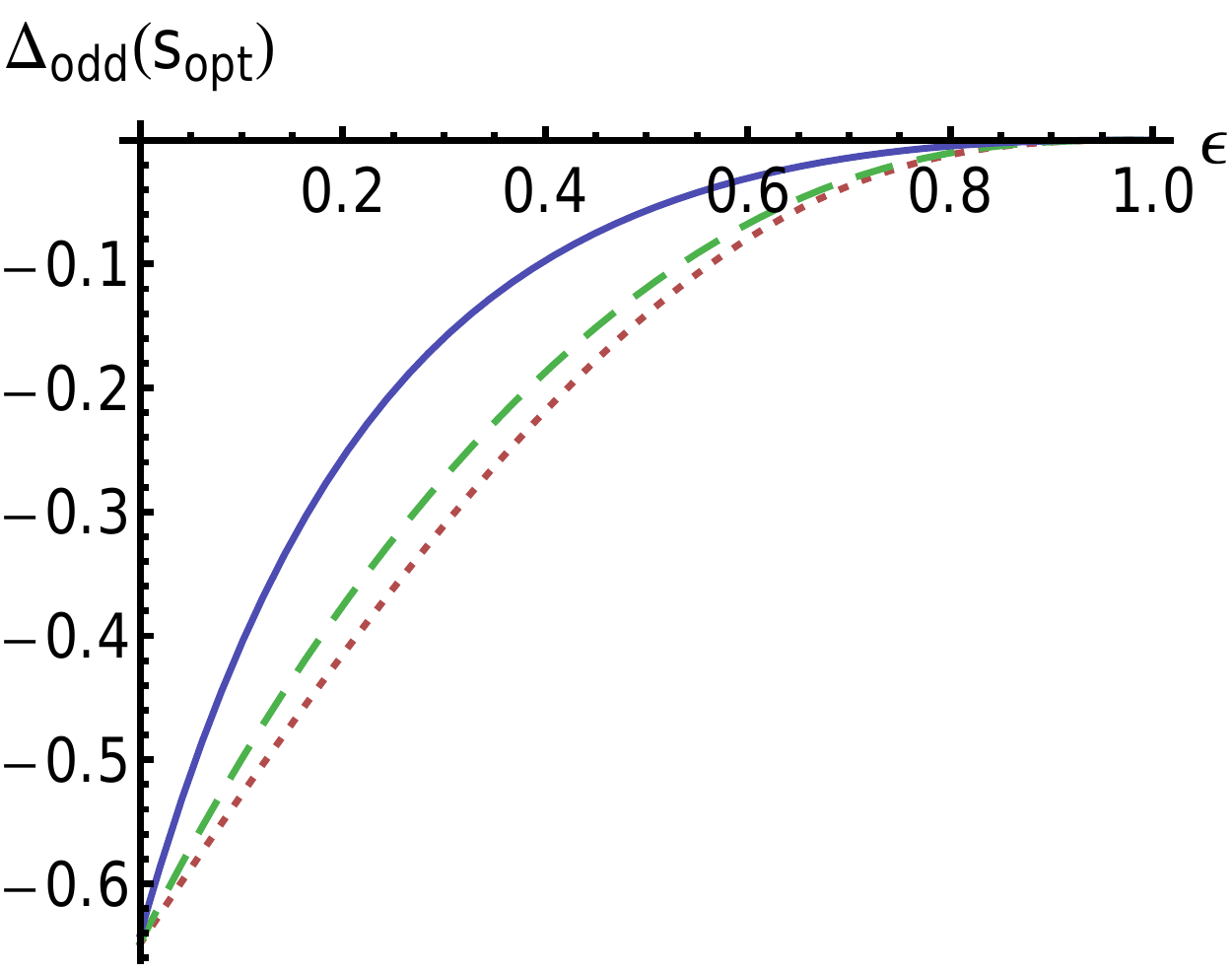}
\includegraphics[width=0.4\columnwidth]{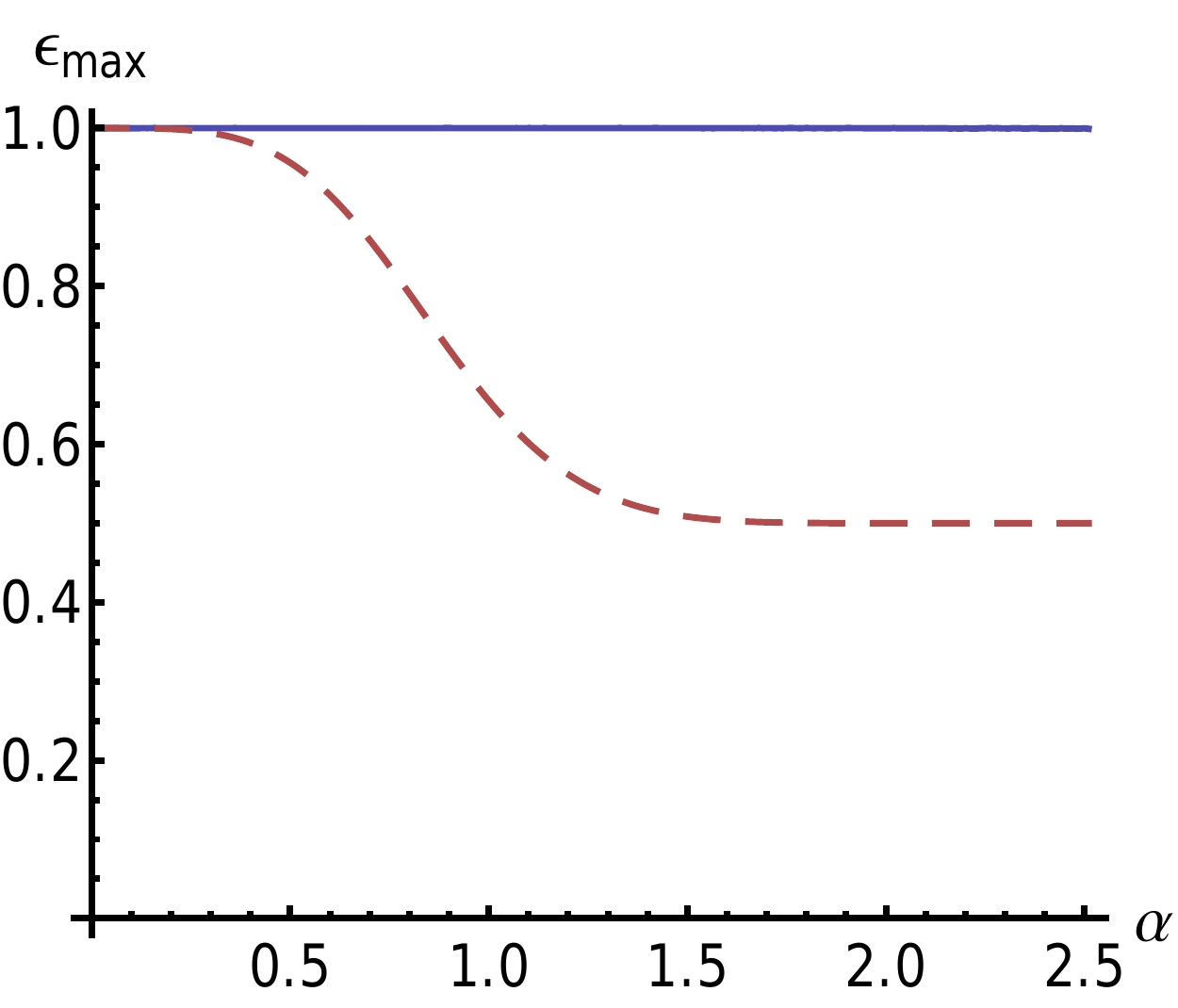}
\caption{(Left) Optimized QNG witness $\Delta_{\sf odd}(s_{\sf opt})$ for the odd cat states evolving in a lossy channel, as a function of $\epsilon$ and for different values of $\alpha$: red-dotted line: $\alpha=0.5$; green-dashed line: $\alpha=1.0$; blue-solid line: $\alpha=1.5$. \\
(Right) Maximum value of the noise parameter $\emax$ such that the optimzed QNG witness $\Delta_{\sf odd}(s_{\sf opt})$ takes negative values, as a function of the coherent states amplitude $\alpha$ (blue-solid line). The dashed-red line corresponds to the maximum value of the noise parameter $\emax$ obtained without considering additional squeezing ({\em i.e.} for $s=0$).
\label{f:DeltaOdd2}}
\end{figure}
The behaviour of the resulting optimized witness $\Delta_{\sf odd}(s_{\sf opt})$ is plotted as a function of $\epsilon$ for different values of $\alpha$ in Fig. \ref{f:DeltaOdd2} (left): for the values of $\alpha$ considered, QNG of odd cat states can be detected by the criterion for all values of $\epsilon$ . As pictured in Fig. \ref{f:DeltaOdd2} (right), it is possible to prove numerically that $\emax$ is equal to one, that is QNG can be detected for all values of noise, for larger values of $\alpha$, also in cases where without additional squeezing, that is by considering $\Delta_{\sf odd}(0)$, $\emax\approx 0.5$.
\subsection{Even cat states}
\begin{figure}[h!]
\centering
\includegraphics[width=0.4\columnwidth]{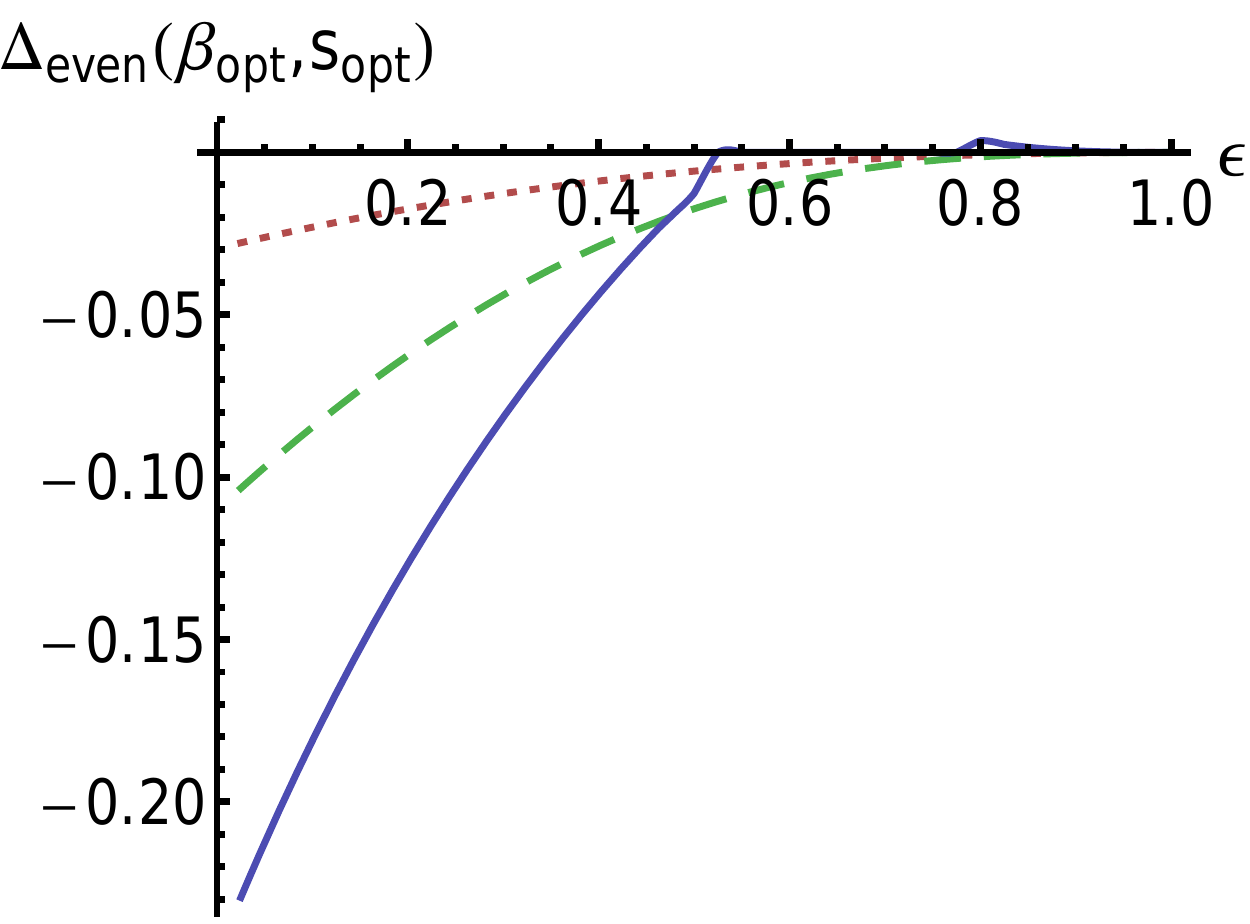}
\includegraphics[width=0.4\columnwidth]{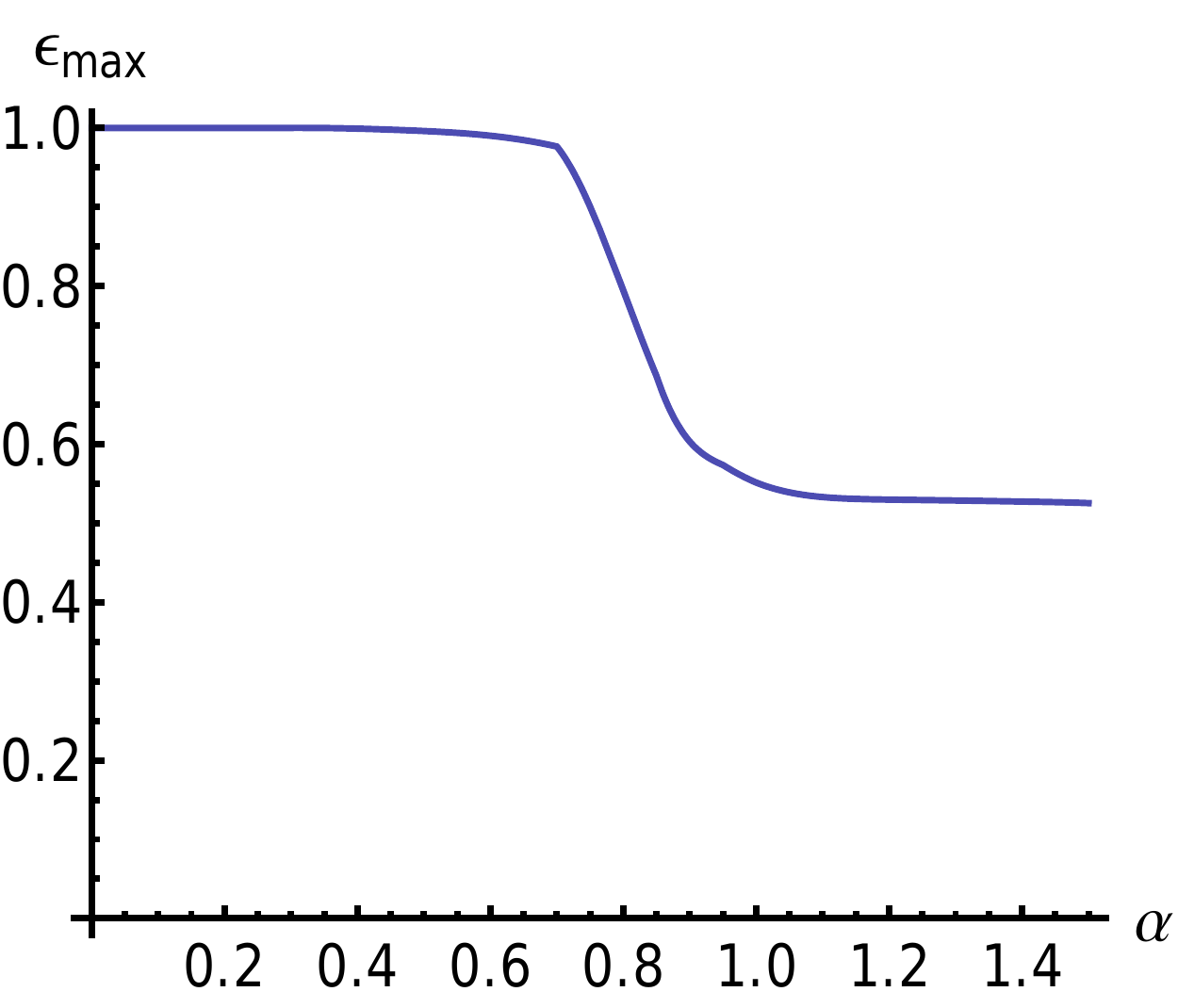}
\caption{(Left) Optimized QNG witness $\Delta_{\sf even}(\beta_{\sf opt},s_{\sf opt})$ for the even cat states evolving in a lossy channel, as a function of $\epsilon$ and for different values of $\alpha$: red-dotted line: $\alpha=0.4$; green-dashed line: $\alpha=0.6$; blue-solid line: $\alpha=1.0$. \\
(Right) Maximum value of the noise parameter $\emax$ such that the optimzed QNG witness $\Delta_{\sf even}(\beta_{\sf opt},s_{\sf opt})$ takes negative values, as a function of the coherent states amplitude $\alpha$. Notice that for $\beta=0$ and $s=0$, one would obtain $\emax=0$ for all values of $\alpha$.
\label{f:DeltaEven2}}
\end{figure}

We now consider the problem of detecting QNG for even cat states $|\psi_{\sf even}\rangle$ evolving in the lossy channel $\E$. By inspecting the plot in Fig. \ref{f:Wigner} (right), we notice that the Wigner function of the initial pure state is squeezed and that its minimum is along the $P$ quadrature axis. As a consequence we will consider a combination of displacement and squeezing operations, in order to construct the following QNG witness 
\begin{eqnarray}
\Delta_{\sf even}(\beta,s) &= \Delta[\E(|\psi_{\sf even}\rangle\langle\psi_{\sf even}|, D(i\beta) S(s)] \:.
\label{eq:Delta2}
\end{eqnarray}
The average photon number $\bar{n}^{\sf (even)}_{\beta,s}$ to be used in the calculation of $\Delta_{\sf even}(\beta,s)$ reads
\begin{equation}
\bar{n}^{\sf (even)}_{\beta,s} = (\mu_s^2 + \nu_s^2) \bar{n}_\epsilon + 2\mu_s\nu_s \langle a^2\rangle_\epsilon + \nu_s^2 + \beta^2 \:, \label{eq:nevens}
\end{equation}
where in this case the values of $\bar{n}_\epsilon$ and $\langle a^2 \rangle_\epsilon$ can be obtained from Eqs. (\ref{eq:nepsilon}) and (\ref{eq:n0}) by setting $\xi=1$.\\
We will look for the optimal values $\{ \beta_{\sf opt}, s_{\sf opt}\}$ which minimize $\Delta_{\sf even}(\beta,s)$ for given values of the noise parameter $\epsilon$ and coherent states amplitude $\alpha$. Unfortunately, in this case an analytical approach cannot be pursued, since the displacement operation changes both the value of the Wigner function in the origin of the evolved state and the average photon number in Eq. (\ref{eq:nevens}). As a consequence the optimal values will be obtained numerically for each couple of values of $\epsilon$ and $\alpha$. 
%
%
The corresponding optimized QNG witness $\Delta_{\sf even}(\beta_{\sf opt},s_{\sf opt})$ has been evaluated and plotted in Fig. \ref{f:DeltaEven2} (left). We can clearly observe that, thanks to the additional Gaussian operations, we are able to detect QNG for non-trivial values of the noise parameter, that is for $\epsilon>0.5$. The maximum value of the noise parameter $\emax$ for which we observe negative values of $\Delta_{\sf even}$ has been obtained numerically and it is plotted in Fig. \ref{f:DeltaEven2} (right). For small values of the coherent states amplitude $\alpha$, one can detect QNG for practically all the possible values of noise:
 in particular for $\alpha \leq 0.1$ we have $\emax \approx1$ (up to numerical precision), and for $\alpha< 0.6$ we still have $\emax > 0.99$. Unfortunately, by further increasing the amplitude up to $\alpha=1$, the witness performances are drastically reduced and $\emax$ approaches its limiting value $\emax \approx 0.5$. Notice that if we do not consider additional operations, that is by setting $\beta=0$ and $s=0$, one obtains $\emax=0$ for all values of $\alpha$, that is QNG cannot be detected.
\section{Conclusions}
We have applied a recently proposed QNG criterion to Schr\"odinger's cat states evolving in a lossy bosonic channel. We observe, that by optimizing the witness by additional Gaussian operations, one can detect QNG, and thus a QNL process, for non-trivial values of the noise parameter, that is for sever optical loss yielding a positive Wigner function. In particular the criterion works really well for {\em odd} cat states, while it is effective only for small amplitude {\em even} cat states.
\section{Acknowledgments}
MGG acknowledges support from UK EPSRC (EP/I026436/1). TT and MSK acknowledge support from the NPRP 4- 426 554-1-084 from Qatar National Research Fund. SO and MGAP acknowledge support
from MIUR (FIRB ``LiCHIS'' No. RBFR10YQ3H).


\section*{References}
\begin{thebibliography}{10}
\bibitem{tutorialMSK} Kim M S 2008 {\it J. Phys. B: At. Mol. Opt. Phys.}  {\bf 41} 133001
\bibitem{AOP}{Ferraro A, Olivares S and Paris M G A 2005 {\em Gaussian States in Quantum
Information}, (Bibliopolis, Napoli, 2005)}
\bibitem{glauber} Glauber R 1963 {\em Phys. Rev.} {\bf 131} 2766.
\bibitem{geno1} Genoni M G, Paris M G A and Banaszek K 2007 {\em Phys. Rev. A} {\bf 76} 042327
\bibitem{geno2} Genoni M G, Paris M G A and Banaszek K 2008 {\em Phys. Rev. A} {\bf 78} 060303
\bibitem{geno3} Genoni M G and Paris M G A 2010 {\em Phys. Rev. A} {\bf 82} 052341
\bibitem{barbieri} Barbieri M \etal 2010 {\em Phys. Rev. A} {\bf 82} 063833
\bibitem{filip1} Filip R and Mista Jr. L 2011 {\em Phys. Rev. Lett.} {\bf 106} 200401 
\bibitem{qnonGPRA}{Genoni M G, Palma M L, Tufarelli T, Olivares S, Kim M S and Paris M G A 2013
{\em Phys. Rev. A} {\bf 87} 062104}
\bibitem{filip2} Jezv‡ek M \etal 2011 {\em Phys. Rev. Lett.} {\bf 107} 213602 
\bibitem{filip3} Jevzek M \etal 2012 {\em Phys. Rev. A} {\bf 86} 043813
\bibitem{cats1} Ourjoumtsev A, Tualle-Brouri R, Laurat J and Grangier P 2006 
{\em Science} {\bf 312} 83
\bibitem{cats2} Wakui K, Takahashi H, Furusa A and Sasaki M 2007 {\em Opt. Express} 
{\bf 15} 3568
\bibitem{cats3} Neergaard-Nielsen J S, \etal
2006 {\em Phys. Rev. Lett.} {\bf 97} 083604
\bibitem{paavola} Paavola J \etal 2011 {\em Phys. Rev. A} {\bf 84} 012121
\end{thebibliography}
\end{document}